# Ultra Low Energy Analog Image Processing Using Spin Based Neurons

[1]Mrigank Sharad, [2]Charles Augustine, [1]Georgios Panagopoulos, [1]Kaushik Roy

Department of Electrical and Computer Engineering, Purdue University, West Lafayette, IN, USA
[2]Circuit Research Lab, Intel labs, Intel Corporation, Hillsboro, OR, US
msharad@.purdue.edu, charles.augustine@intel.com, gpanagop@purdue.edu, kaushik@purdue.edu

*Abstract*- **In this work we present an ultra low energy, 'on-sensor' image processing architecture, based on cellular array of spintronic neurons. The 'neuron' constitutes of a lateral spin valve (LSV) with multiple input magnets, connected to an output magnet, using metal channels. The low resistance, magneto-metallic neurons operate at a small terminal of ~20mV, while performing analog computation upon photo sensor inputs. The static current-flow across the device terminals is limited to small periods, corresponding to magnet switching time, and, is determined by a low duty-cycle system-clock. Thus, the energy-cost of analog mode processing, inevitable in most image sensing applications, is reduced and made comparable to that of dynamic and leakage power consumption in peripheral CMOS units. Performance of the proposed architecture for some common image sensing and processing applications like, feature extraction, halftone compression and digitization, have been obtained through physics based device simulation framework, coupled with SPICE. Results indicate that the proposed design scheme can achieve ~100X reduction in computation energy, as compared to the state of art CMOS designs, that are based on conventional mixed-signal image acquisition and processing schemes.**

## I. INTRODUCTION

Recent years have seen a sustained thrust towards integration of increasingly complex image processing functionality on CMOS photo-sensor arrays, for real-time [1, 2, 5, 7], and high speed imaging applications [3, 8]. In almost all such designs, analog-mode computation is inevitably present, in some form or the other. As a result, in most of the 'on-sensor' image processing architectures, analog units like, comparators, current mirrors and ADC's present in each of the pixels, account for most of the computation energy per frame [40-42]. Apart from on-sensor image quantization, emerging high-performance vision-IC's incorporate several low and middle-level image processing tasks, like, averaging, edge detection, motion detection, and object tracking. Such operations often involve cellular processing of pixels values, based on near-neighborhood computation using cellular neural networks (CNN) [9, 10]. Analog, rather than digital-mode, processing has been argued to be suitable for such computations, owing to compactness of the analog modules [4]. However, most of the recent designs reported large power consumption, especially, for high frame rates, resulting, mainly, from the analog processing elements [1-10]. Moreover, design of analog circuits becomes increasingly more challenging at scaled technology nodes, and, power consumption for the same performance can increase for several common analog circuits [11]. Therefore, conventional analog designs may not be the most suitable candidates for high integration-density and low cost signal processing hardware, needed for the fast evolving imaging technology. Hence, it is desirable to look for alternate device technologies that can carry out the analog processing tasks at low energy and low real-estate cost.

In this work, we explore the application of spintronic devices, based on lateral spin valve (LSV), in image acquisition and processing hardware. An LSV constitutes of *nano-magnets* connected through non-magnetic metal channels (fig. 1a) [12-19]. The *nano-magnets* in an LSV can interact and undergo spin transfer torque (STT) induced switching. Logic schemes based on LSV have been explored by several authors [14-18].

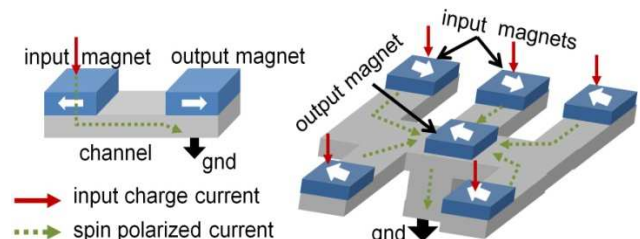

Fig. 1(a) Lateral spin valve (b) five input majority gate

Analog characteristics of the current mode switching scheme employed in an LSV, make it suitable for non-Boolean computation like majority evaluation (fig. 1b), and, enable it to handle analog inputs. In [21, 22, 43] it was shown that, a spin majority gate with adjustable spin injection strength of the input magnets and appropriate clocking of the output magnet mimics neuron operation. Such a magneto-metallic device can operate at a small terminal voltage (~20 mV) and can be employed in low power analog computation.

In this work we present the application of such a 'spintronic neuron', in an 'on-sensor' image processing architecture. We show that, the spintronic neurons can be integrated with CMOS transistors to arrive at spin-CMOS hybrid processors (PE). In such a PE, the analog-mode computation can be carried out with the help of the neurons, at ultra low energy cost. Apart from ultra low voltage operation, the fast switching of the neuron-magnets also help in reducing the computation energy. This comes from a clock synchronized computation scheme, where the static current flows only for a period close to *nano-magnet* switching time, which can be very small, as compared to the highest frame rates of practical interest.

Rest of the paper is organized as follows. Device structure for the spintronic neuron is described in section 2. Section 3 briefly introduces the concept of cellular neural network. Circuit level integration of the neuron device to realize the CNN functionality is presented in section 4. Section 5 presents simulation results for some common image processing applications. Section 6 briefly describes the simulation framework used in this work. In section 7 we discuss the performance and prospects of the proposed scheme. Finally, section 8 concludes the paper.

## II SPIN BASED NEURON MODEL

In this section we introduce the spintronic neuron model. The basic device operation for LSV structures, with decoupled read and write paths, is first explained. We then describe the functionality of the neuron device that is based on these structures.

### A. Lateral spin valves (LSV) with decoupled read and write paths

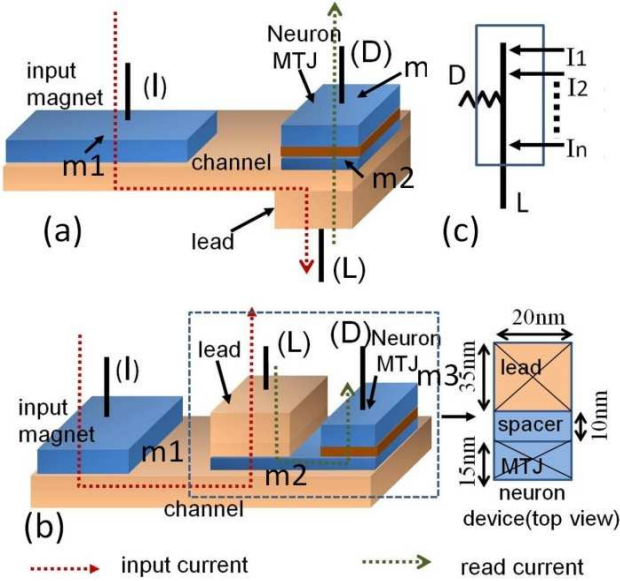

Fig. 2 (a) LSV with non-local STT switching, (b) Read-write decoupled LSV with local STT switching. (c) three terminal schematic model of LSV with decoupled read-write

Two different LSV structures with decoupled read and write paths are shown in fig.2.

The device in fig.2a employs 'non-local' spin torque for *nano-magnet* switching [12, 13]. The figure shows a high-polarization ($P$) input magnet $m_1$ which acts as a spin injector and a low-polarity output magnet $m_2$, which forms a magnetic tunnel junction (MTJ) with a fixed magnet $m_3$. Charge current injected into the channel through $m_1$ gets spin polarized according to the polarity of $m_1$. Spin-polarized charge current is modeled as a four component quantity, one charge component $I_c$, and three spin components ($I_x, I_y, I_z$) [14, 15]. The charge component flows into the lead. A portion of the spin component however, is absorbed by the low-$P$ interface of $m_2$ and exerts spin torque on it. Rest of the spin component is lost into the lead. Owing to the separation of the spin component, responsible for spin-torque, from the charge-current flow, this scheme is regarded as 'non-local' spin transfer torque (STT) switching [12, 13]. Experimentally, ~20% efficiency for non-local spin injection (ratio of spin absorbed by the output magnet to the spin current injected into the channel) in LSV has been demonstrated [12, 15]. However, simulation based analysis shows that, this efficiency can be further enhanced by geometrical optimization of the device structure [18, 21, 22].

The second LSV structure, shown in fig. 2b, employs a relatively larger size for the output magnet $m_2$, in order to achieve decoupled read and write. Around 60% of its top area (35nm x 20nm) is occupied by a metal lead through which the switching current flows, whereas, a smaller portion (15nm x 20nm) is used as a read-port that constitutes of an MTJ. Note that, although, the input current flows only though a part of $m_2$, its small dimension (60x20x1) ensures mono-domain behavior and switching of the entire magnet is achieved. But, the switching current required, for the same switching time, is almost twice as compared to the case, when the extended area of the magnet, forming the read port is absent. Owing to the direct current injection into output magnet, this structure can, however, achieve higher spin-injection efficiency. For a high polarity interface of $m_1$ (P~0.9) and a low polarity interface of $m_2$ (P~0.1), almost ~90% injection can be obtained provided the channel length is within the spin diffusion length of the channel material (~1μm for copper) [15].

Both the LSV structures can be represented as a three-port unit (fig. 2c). The input port(s), $I$, the lead terminal, $L$, and the detection terminal $D$. The input currents flow between the terminals $I$ and $L$, i.e., through a low resistance, metallic path. Hence, a small terminal voltage across these two terminals can drive the required switching current. The terminal $D$ is used to detect the state of the output magnet, $m_2$, without injecting static current into the high resistance tunneling barrier (using dynamic CMOS latch discussed later).

Next, we show the application of the LSV structures described above in realizing the neuron functionality.

### B. Neuron device

Fig. 3 shows the device structure for neuron based on LSV. It constitutes of an output magnet $m_1$ with MTJ based read-port, and three input magnets, $m_2$-$m_4$. The two anti-parallel, stable polarization states of a magnet lie along its easy axis (fig. 3). The direction orthogonal to the easy axis is an unstable polarization state for the magnet and is referred as its hard-axis [14, 16]. The two input magnets, $m_2$ and $m_3$, possess anti-parallel spin-polarization, and, have their easy-axis parallel to that of $m_1$. The preset magnet $m_4$ shown in fig. 3, however, has its easy-axis orthogonal to that of $m_1$, and is used to implement current-mode Bennett-clocking [14, 16]. A current pulse input through $m_4$ presets the output magnet, $m_1$, along its hard axis (fig. 4). The preset pulse is overlapped with the synchronous input current pulses received through the magnets, $m_2$ and $m_3$. After removal of the preset pulse, $m_1$ switches back to its easy axis, which is parallel to that of $m_2$ and $m_3$. The final spin polarity of $m_1$ depends upon the difference $\Delta I$, between the spin polarized charge current inputs through $m_2$ and $m_3$, (fig. 1b). Hard-axis, being an unstable state for $m_1$, even a small value of $\Delta I$, effects deterministic easy-axis restoration. Note that, the lower limit on $\Delta I$ for deterministic switching is imposed by the thermal noise in the output magnet [14, 21]. Thus, the neuron device essentially acts as an ultra low voltage spin-mode comparator.

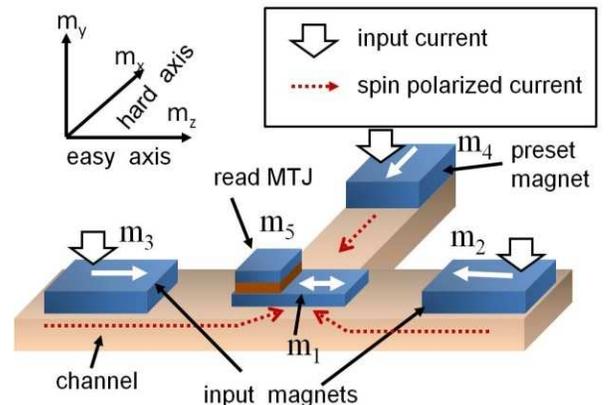

Fig. 3 Spintronic neuron with two complementary inputs.

In [22] we showed that the spin based neuron discussed above can be integrated with programmable conductive elements like memristors [49], to realize low-power computational neural networks. In this work we show that with the help of CMOS transistors, operating in deep-triode region, the device can be used to implement ultra low power processer (PE) for CNN based image processing architecture.

In the next section we introduce the CNN paradigm. Following this we describe circuit schemes employed to realize the CNN functionality with the neuron model described above.

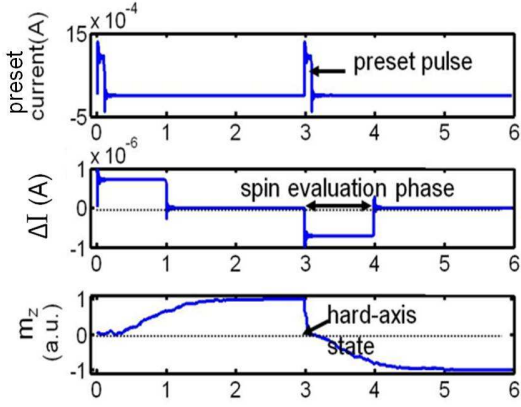

Fig. 4 Timing waveform for the proposed neuron model

## II. CELLULAR NEURAL NETWORK : MATHEMATICAL MODEL

Cellular neural network (CNN) can be regarded as a fusion of artificial neural network (ANN) and cellular automata [4, 9, 10, 27-29]. It borrows the basic information processing functionality, i.e., the 'integrate and fire' operation upon weighted inputs, from neural networks. The concept of computation based on neighborhood influence, on the other hand, is akin to cellular automata. This class of computation has been found to be highly suitable for several image processing applications, which essentially involve processing of pixel neighborhoods in a parallel fashion.

Fig. 5 shows a cellular neural network array with 3x3 rectangular neighborhoods. Each cell is connected to its eight surrounding neighbors through a 3x3 feedback-weight template $A$. $A(0,0)$ denotes the self feedback term. The feed-forward template of a cell, $B$ (or the input-weight template), determines the connectivity to the neighborhood inputs. In a CNN, each neuron performs 'integrate and fire' operation upon the weighted combination of its neighborhood inputs and outputs in a recursive manner.

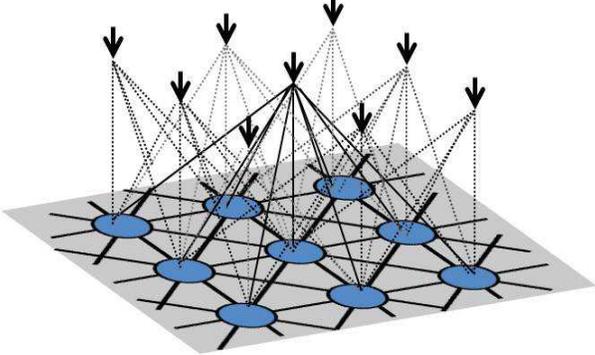

Fig.5. CNN architecture with 3x3 neighbourhood connectvity

The standard expression for a CNN cell state is given by eq. 1 [27].

$$C\frac{dx_{ij}(t)}{dt} = -x_{ij}(t) + \sum_{(k,l)\in N(i,j)} A(i,j;k,l).y_{kl}(t) + \sum_{(k,l)\in N(i,j)} B(i,j;k,l).u_{kl}(t) + z(i,j) \quad (1)$$

Where, $x(t)$ is the cell state at time $t$, $A$ and $B$ are the feedback and feedforward template defined above, $u(t)$ is the input to cell from its 3x3 neighborhood $N$ and $z$ is the cell-bias. The cell output is denoted by $y(t)$ which is related to the cell state $x(t)$ with a non-linear transfer-function. Time domain dicretization of the CNN state equation leads to eq. 2 [28].

$$x_{ij}(k) = \sum_{(k,l)\in N(i,j)} A(i,j;k,l).y_{kl}(k) + \sum_{(k,l)\in N(i,j)} B(i,j;k,l).u_{kl}(k) + z(i,j) \quad (2)$$

Discrete time CNN (DTCNN) employs a step transfer function given by eq. 3.

$$y_{ij}(k) = f'(x_{ij}(k-1)) = \begin{cases} +1 & if \ x_{ij}(k-1)>0 \\ -1 & if \ x_{ij}(k-1)<0 \end{cases} \quad (3)$$

Although, in literature, DTCNN templates for image processing applications have been generally obtianed for bipolar output levels, the network functionality is preserved for any two values for the binary states. Hence, DTCNN templates obtained for bipolar transfer function given by $f(x)$, in general, can be used for a step transfer function with arbitrary binary levels. For instance, the effect of a non-zero offset in $f(x)$ can be included in eq. 2 by adding an offset matrix, $U$, with all elements equal to the offset value (eq. 4).

$$x_{ij}(k) = \sum_{(k,l)\in N(i,j)} A(i,j;k,l)(y_{kl}(k)+U_{3X3}) + \sum_{(k,l)\in N(i,j)} B(i,j;k,l).(u_{kl}(k)+U_{3X3}) + z'(i,j) \quad (4)$$

In order for the cell state to remain unchanged, we only need to update the cell-bias $z$, as in eq. 5.

$$z'(i,j) = z(i,j) - \sum_{(k,l)\in N(i,j)} A(i,j;k,l)U_{3X3} - \sum_{(k,l)\in N(i,j)} B(i,j;k,l).U_{3X3} \quad (5)$$

Unipolar inputs and unipolar neuron transfer-function reduces the complexity of hardware realization significantly. Hence, we chose unipolar binary states for the neurons in this work, resulting in the neuron transfer-function given by eq. 6.

$$y_{ij}(k) = f'(x_{ij}(k-1)) = \begin{cases} 1 & if \ x_{ij}(k-1)>0 \\ 0 & if \ x_{ij}(k-1)<0 \end{cases} \quad (6)$$

Application of a step transfer function limits the value of a cell output $y(i,j)$ to binary levels of $f'(x)$. The input $u(i,j)$, however, can assume continuous values corresponding to the range of pixel intensity.

In the spin-CMOS hybrid PE proposed in this work, the two input magnets ($m_2$ and $m_3$ in fig. 3) of the neuron device shown in fig. 3 are used to realize the inter-neuron connectivity through $A$ and $B$ templates respectively. All the neighboring outputs $y(i,j)$ (/inputs $u(i,j)$ ) linked to a neuron with positive $A(i,j)$'s (/$B(i,j)$'s) connect to one of the inputs, say $m_2$, whereas, those, assosiated with negative terms in the template matrices, connect to the other input $m_3$. The circuit techniques employed to realize a DTCNN processor (PE) with the spintronic neuron is described in the next.

## III. DTCNN ARCHITECTURE WITH SPINTRONIC NEURONS

In this section we describe the design of spin-CMOS hybrid PE that implements the DTCNN funtionality for on-sensor image processing. The inputs signal $u(i,j)$ for a cell, is the associated photo-sensor input. Transistors of weighted dimensions are used as deep-triode region current sourses (DTCS), to implement $A$ and $B$ templates. The neuron in a PE, receives sensor input signals and outputs of its neighboring PE's through the DTCS's in the form of charge current. The current mode signals combine in the metal channel of the neuron, where the Bennett clocking of the output magnet realizes, eq. 3. A dynamic-CMOS detection unit however, converts the bipolar spin information pertaining to the state of the neuron-magnet, into unipolar voltage-level. Hence, the final PE output is given by eq. 6. The circuit operation corresponding to these step are described in the following paragraphs.

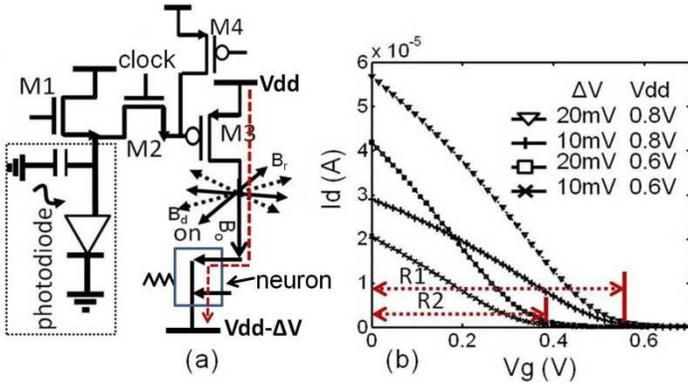

Fig. 6 (a) Circuit for *B*-template realization (b) deep-triode region characteristics of the DTCS transistor $M_3$ driven by the sampled photo-sensor voltage.

Fig. 6 shows a photodiode that converts the illumination intensity received at a pixel into a voltage signal. The transistor $M_1$ first presets the photodiode capacitance to $Vdd-V_t$, where $Vdd$ is the supply voltage and $V_t$ is the threshold voltage of the transistor. The capacitance is then discharged by the photodiode current, rate of discharge being proportional to the incident illumination intensity [7]. At the end of discharge period of a fixed duration, the transistor $M_2$ samples the photodiode voltage. The sampled voltage at the gate of $M_3$ ranges from $Vdd-V_t$ to 0V, corresponding to the illumination intensity at the pixel. $M_3$ supplies input current to the neurons located in the 3x3 neighborhood of the pixel through separate and weighted fingers, with dimensions corresponding to the elements of the *B* template. A second DC level $Vdd-\Delta V$ is used in the design, in order to exploit the low-voltage operation of the spintronic neurons. It connects to the lead terminal of the neurons as shown in fig. 6a. The current supplied by $M_3$ therefore, flows through a small terminal voltage $\Delta V$, which can be of the order of ~10mV. Note that, since the resistance of $M_3$ is significantly higher than that of the magneto-metallic neurons, it accounts for most of the $\Delta V$-voltage drop. Fig. 6b shows that the output current of $M3$ is a fairly linear function of the sampled gate voltage for the deep-triode region operation.

Fig. 7 shows the circuit scheme used to realize the *A*-template. The corresponding simulation waveforms are shown in fig. 8. When the clock is low, output of the dynamic-CMOS latch is precharged to $Vdd$. The latch is activated at the positive edge of the clock signal. The two load branches of the latch are connected to the detection terminal, *D*, of the neuron and a reference MTJ respectively. The latch compares the difference between the effective resistances in its two load branches through a transient discharge current. It drives negligible static current into the high resistance neuron-MTJ stack. For the anti-parallel state of the neuron-MTJ ( which can be regarded as the 'firing state'), the latch drives the DTCS transistor $M_s$ shown in the figure. $M_s$, in turn, supplies current to the neighbouring neurons through separate weighted fingers corresponding to the *A* template. After a time delay that is sufficient for the latch to evaluate and settle to its final value, the neuron device receives the preset current through a clock driven DTST (fig. 8). Note that, a delayed preset pulse with respect to the clock edge ensures that the latch evaluates correctly according to the neuron-MTJ state stored in the previous evaluation cycle. Once evaluated, the latch can not change its state until it is precharged again, despite the flipping of the neuron MTJ. At the positive edge of the clock, the latches in all the PE's evaluate simeltaneously and conditionally drive their respective DTCS outputs. Hence, a neuron recieves input currents from its neighbors, during the period when the clock is high.

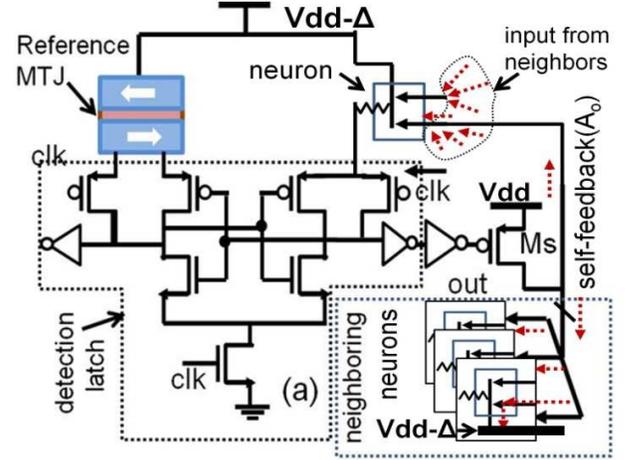

Fig. 7 CMOS detection unit senes the state of the neuron magnet and transmits current mode signal to the neighboring neurons through a deep triode current source transistor.

As soon as the preset signal goes low, the neuron magnet settles to one of its stable states, depending upon the overal spin current received through its inputs. Thus, the recursive operation of DTCNN PE, given by eq. 2 is realized by the application of an appropriate clocking scheme. Note that, the current supplied by the DTCS outputs of the latches also flow across the two supply levels, $Vdd$ and $Vdd-\Delta V$, as shown in fig. 7.

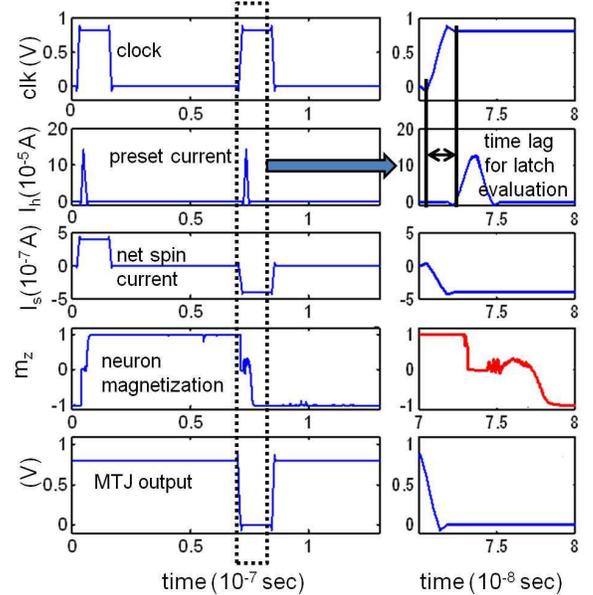

Fig. 8 Simulation waveform for DTCNN operation of the spin-CMOS hybrid PE.

Fig. 9a shows the layout for the CMOS circuitry employed in the spin-CMOS hybrid PE. It shows that a major portion of the PE area is occupied by the triode-region source-transistors ($M_3$ in fig.6a and $M_s$ in fig. 7 ). As mentioned earlier, in order to realize non-overlapping inter-neuron connectivity, we employed separate fingers in the source transistors. Moreover, a matched layout of the fingers was considered. Fig. 9b shows the values of *A* and *B* templates for two common applications, halftoning and edge detection ( results for which have been given in sec. 5). As mentioned before, for an application specific design, the fingers of DTCS's are weighted according to the templates. In the simplest case, for a given connectivity, number of fingers equal to the weight

(matrix element) magnitude can be chosen. The sign of the weight, determines the connectivity, to one of the complementary input of the corresponding neuron.

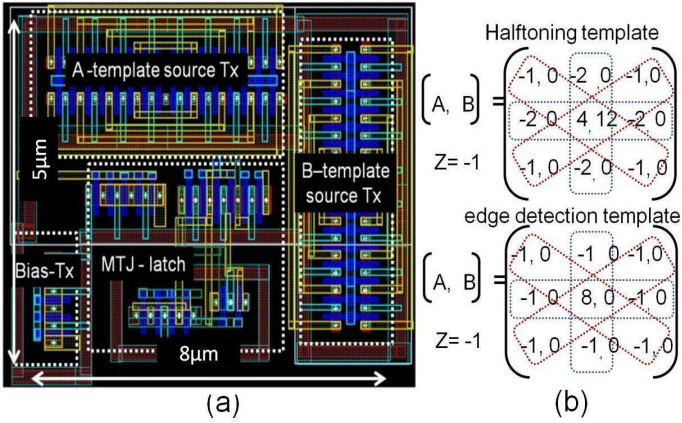

Fig. 9 (a) Layout of the CMOS circuit (90 nm technology) in the PE showing that the source transistors occupy larger portion of the PE area. (b) DTCNN templates for edge detection and halftoning

As discussed before, application of current mode Bennett-clocking reduces the required amount of current injection for a neuron, per-input, to few microamperes. Hence, the multi-finger DTCS transistors can supply the required current even at a small terminal voltage $\Delta V$. Therefore, two DC supply levels separated by a difference of ~20mV can be chosen. This achieves reduced static power consumption for current-mode inter-neuron signalling.

As long as input currents of the neurons are large enough to overcome the impact of thermal noise in the neuron-magnet, the precision of computation achievable, with the proposed scheme, is limited, mainly, by the supply noise. As the accuracy of on-chip DC supply regulation, in the state of art technology is limited to ~0.1% [44], high precicion imaging applications may seem out of scope of the proposed design. However, the use of dual supply rails proposed in this work may significantly compensate this disadvantage. Differential supply lines can significantly mitigate the impact of the noise sources, that lead to common-mode fluctuations. Hence a thorough modelling and analysis of this effect needs to be considered, in order to assess the noise tolerance of the proposed scheme. In the present work, we have included the effect of supply and process variations, and we discuss these in the next section on simulation framework.

## IV. SIMULATION FRAMEWORK

The device simulation used in this work is based on self-consistent solution of spin-transport and Landau-Lifshitz-Gilbert equation (LLG) for the neuron device, and, has been benchmarked with experimental data on spin valves [14-16]. Effective noise field was included in LLG (based on stochastic LLG [14]) in order to account for the thermal noise on device performance [14]. Simulation of MTJ is based on self-consistent solution of LLG and spin transport. Fig. 10 depicts the device-circuit co-simulation framework employed in this work to assess the system level performance. Behavioral model for the neuron device, derived from the physics based equations, was used for simulating large image processing arrays. CMOS design parameters like, voltage levels, clock duty cycle, required current injection and the associated transistor sizing etc, were determined on the basis of device characteristics. On the other hand, state of art circuit limitations were considered in determining appropriate operating conditions for the spin device.

In order to account for the CMOS process variation upon system performance 15% 3σ variations in transistor threshold was considered. Independent noise sources (with uniform distribution) were added to the two supply lines corresponding to 0.1% peak-to-peak voltage fluctuation. The effect of these variations have been shown in the next section.

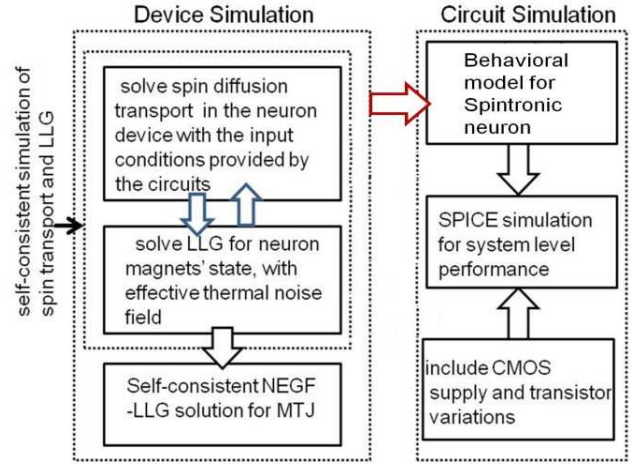

Fig. 10 Device-circuit co-simulation framework used in this work

## V. APPLICATION SIMULATION

In the following sub-sections we present simulation results for some common image processing applications like edge detection, halftoning and digitization.

### A. Feature extraction

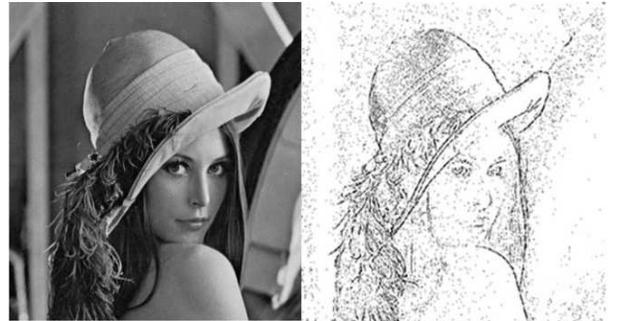

Fig. 11. Result of edge detection from a grey-scale image

Edge detection (fig. 11) is one of the most common image processing steps, applied in vision applications [30, 31]. As an example, motion detection (fig. 12) employs comparison between the edge maps of a still background, sampled one after the other. This can be achieved by employing extra storage registers per PE to store a sequence of edge maps.

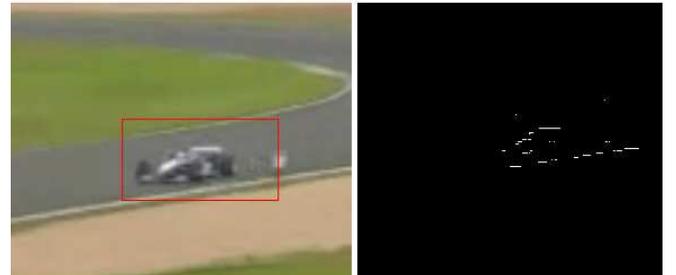

Fig. 12 Motion detection on the basis of temporal difference in edge maps.

### B. Halftone compression and sensisng

Halftoning is a process in which a grey scale image is recorded as (or compressed into) a binary image, with just two levels, in a way such that important details in the image are preserved [33]. Several algorithms for decompressing halftone images have been proposed in literature [32]. This technique can be used for sensing, storing and transmitting images in bandwidth limited sytems. Simulation result for halftoning of a statellite image is shown in fig. 13. Fig. 14 shows the halftoned image of Lenna along with the effect of reduction in *ΔV* upon the halftone output. With decreasing *ΔV* the effect of noise becomes increasingly more prominent.

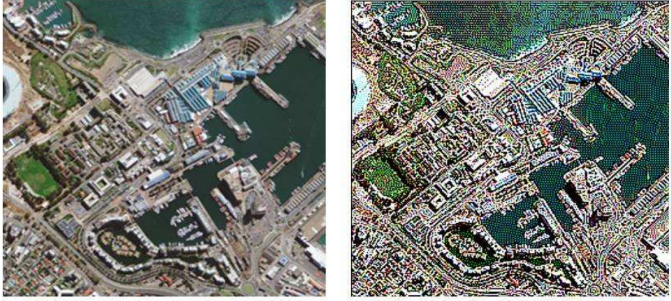

Fig. 13 Simulation results for halftoned image of a satellite picture.

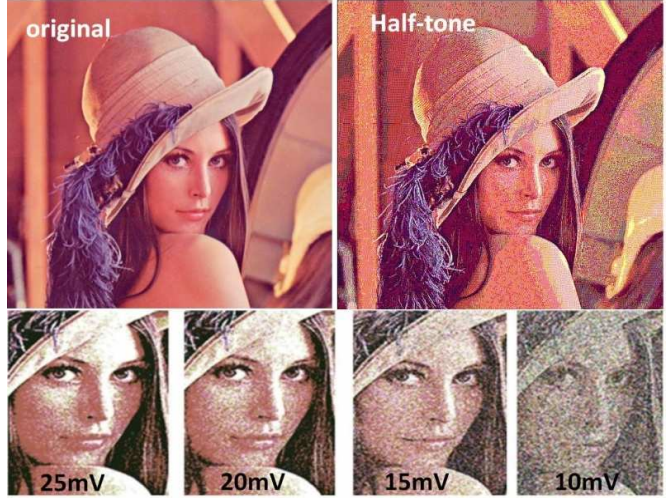

Fig. 14 (a) Halftone of Lenna (b) effect of reduction in *ΔV* upon the output, with 0.1% supply noise and constant DTCS width (i.e. reducing current and increasing % noise)

### C. Digitization

Successive-approximation-register (SAR) analog-to-digital converter (ADC) is one of the most common data converters employed for on-sensor image quantization (fig 15a) [33]. The data conversion algorithm employed in an SAR-ADC can be explained as follows. To begin the conversion, the approximation register is initialized to the midscale (i.e., all but the most significant bit (MSB) is set to 0. At every cycle a digital to analog converter (DAC) produces an analog level corresponding to the digital value stored in the register, and, a comparator compares it with the input sample. If the comparator output is high, the current bit (MSB) remains high, else, it is turned low and the next bit is turned high. The process is repeated for all the bits. At the end of conversion, the SAR stores the digitized value for the pixel intensity, which can be read out in a column-wise manner from the sensor array.

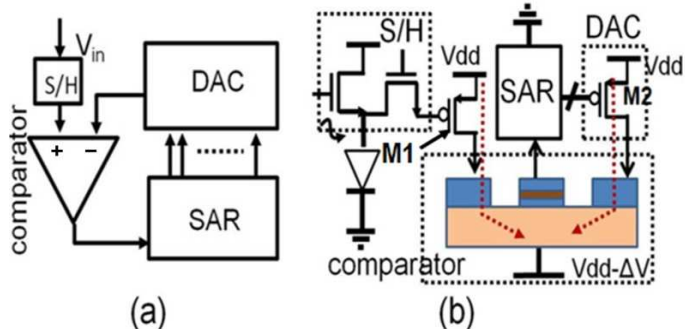

Fig. 15(a) SAR ADC block diagram (b) compact and low power SAR ADC using spintronic neuron.

In a cicuit implementation of SAR-ADC, most of the power consumption results form the comparator and the DAC units [33]. The SAR unit consists of a bank of CMOS latches and a simple control logic, which consumes negligible power as compared to the analog units.

As the SAR-ADC essentially employs recursive evaluation, akin to the CNN equation, the PE circuit decribed in the previous section can be easily extended to realize a compact and low power *N*-bit SAR-ADC. In the schematic diagram for the proposed ADC, shown in fig. 15b, the DTCS $M_1$ converts the sampled output of the photo sensor into a current signal, that is injected into one of the inputs of a three input neuron. The SAR simply consists of a bank of *N* CMOS latches, which in turn drive *N* different fingers of the DTCS *M2*. The multiple fingers of *M2* are binary weighted (for *N*=8, the weakest transistor having 4X minimum length and minimum width, and largest transitor having 8 fingers with 4X minimum width and minimum length) and hence, it acts as a compact DAC and injects current into the second complementary input of the neuron. Current mode Bennett-clocking of the neuron, using the third input (a preset magnet, not shown in fig. 15b), at the beginning of each conversion stage, realizes the comparator operation. Note that, in the proposed ADC design, the analog computation current flows across the two supply levels, i.e., across a small terminal voltage *ΔV*, thereby, resulting in small power consumption. Moreover, in each frame, the current flow is restricted to the small period of conversion just after the data is sampled.

Fig. 16 shows the simulation results for an 8-bit SAR-ADC based on the proposed scheme. Degradation in image quality due to supply noise can be perceived. Note that, in this work we have not considered any coupling between the two supply levels and independent noise sources have been used in simulation. Hence a thorough analysis of the proposed differential supply scheme would be needed to assess the computation precision, achievable by the proposed hardware.

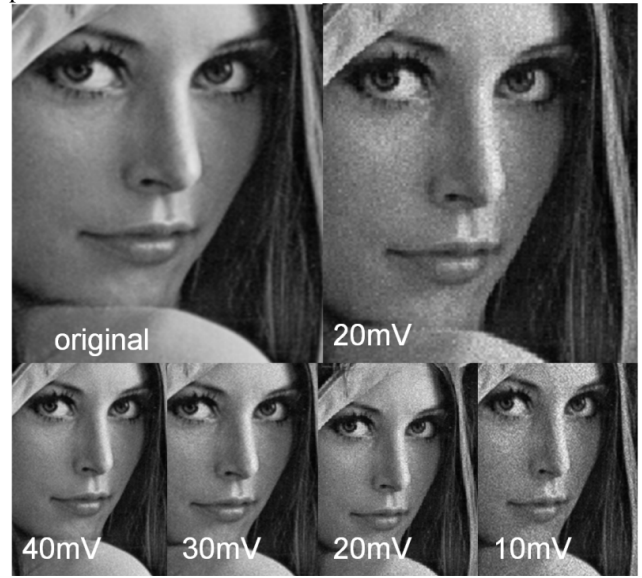

Fig. 16 Simulation result of spin-CMOS hybrid 8 bit-SAR-ADC and the effect of lowering *ΔV* upon the output, with 0.1% supply noise and constant current (by increasing DTCS widths).

## VI. DESIGN PERFORMANCE

Fig. 17 depicts the architecture for on-sensor image processing [7]. Such a design employs PE's integrated on each of the photo-cell. The output of the photo-detectors are directly processed by the PE's and the result is read out column-wise.

In such an architecture, the total energy dissipation per-input frame can be expressed as the sum of computation energy ($E_{comp}$), the read-out energy ($E_{read}$) and the energy that is wasted in the form of leakage current ($E_{leak}$).

$$E_{tot} = E_{comp} + E_{read} + E_{leakage} \quad (7)$$

In this work, $E_{comp}$ can be expressed as a sum of neuron-preset-energy, (the energy associated with current mode Bennett-clocking), $E_{preset}$, the energy associated with current mode inter-neuron signaling, $E_{evl}$, and the dynamic switching energy in the PEs', $E_{dynamic}$ (including energy consumption due to clocking). A first order expression for these components can be derived using the design parameters, namely, the two supply levels $Vdd$ and $Vdd$-$\Delta V$, the read-out voltage $V_{read}$, the preset time $T_{pre}$, the evaluation time $I_{evl}$, the effective switched capacitance in a PE, $C_{PE}$, the bit-line capacitance $C_{BL}$, the word-line capacitance $C_{WL}$, number of cells in the array $N \times N$, the switching activity factor, $\alpha$, and the number of iteration required per-frame for a given operation, $M$:

$$E_{comp} = N^2 M (E_{preset} + E_{evaluation} + E_{dynamic})$$
$$= N^2 M (\Delta V T_{pre} I_{pre} + \Delta V T_{evl} I_{evl} + \alpha C_{PE} V_{dd}^2) \quad (8)$$
$$= N^2 M (\Delta V T_{pre} I_{pre} + \Delta V T_{evl} I_{evl} + \alpha C_{PE} V_{dd}^2)$$

The read-out energy, in the case of column-wise read-out can be obtained using the effective bit-line capacitance that is switched to read out $K$ bit data per PE from the entire $N \times N$ frame,

$$E_{read} = KN(N(\alpha' C_{BL} V_{dd} V_{read}) + \alpha' C_{WL} V^2) \quad (9)$$
$$\approx KN^2 (C_{BL} V^2)$$

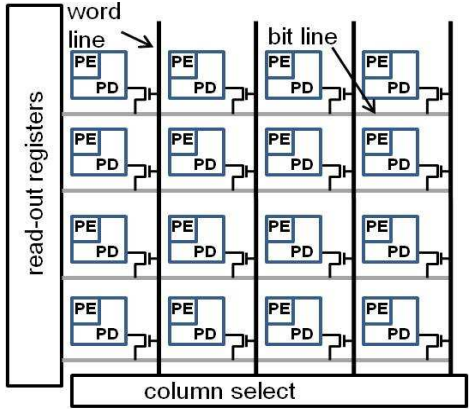

Fig. 17 An on-sensor image processing architecture contains PE's embedded into the pixel locations, and an addressing arrangement for reading out the PE outputs in a column-wise manner.

$E_{leak}$ can be ignored, as there are well known gating techniques that can make the leakage power for the PE's negligibly small during the read-out period.

The results given in table-1, based on the design parameters in table-2 and table-3, indicate that for the proposed architecture, $E_{comp}$ is of the same order as $E_{read}$. Hence, the energy component, related to static power consumption due to analog-mode computation, can become comparable to that associated with dynamic power consumption in the peripheral digital-circuits.

As described earlier, the advantage of using the proposed spin-CMOS hybrid scheme for analog computation comes from two main factors. The first, static current flow across a very small voltage $\Delta V$, and the second, pulsed operation of the spintronic neurons with a narrow pulse-clock. Although, gating of analog modules in low frame rate image processing architectures have been proposed [45], gating of analog circuits for high frame rates can be challenging. Moreover, it might be difficult to gate analog cirucits with a pulse-width of a few nano-seconds, which is possble with the spintronic neurons.

Comparison with on-sensor image processing designs for feature extraction, given in table-IV, shows more than two orders of magnitude improvement in computation energy. Note that, the effect of technology scaling has been included through a mutiplicative factor of $S^2$, where, $S$ is the ratio of the technology scale between the reference design and the presented work (90nm CMOS) [48]. Figure of merit (FOM) is evaluated on the basis of computation energy per frame (as given under table-V),

Table-V compares the performance of the proposed SAR-ADC with some recent CMOS designs. Note that ADC is one of the few analog modules for which power consumption reduces with scaling [11, 48]. Results show that the spin-CMOS hybrid ADC can achieves ~50x low power consumption, as compared to some of the latest designs.

Table-I
Design Performance for 256x256 array

| Frame rate: 10000 fps | $E_{comp}$ | $E_{read}$ | Power |
|---|---|---|---|
| 8-bit quantization | 13nJ | 8nJ | 180μW |
| Edge detect. | 4nJ | 1nJ | 40μW |
| Halfton. | 6nJ | 1nJ | 50μW |

Table-II
Design Parameters (90nm CMOS)

| Vdd | 900mV | $C_{PE}$ | 6fF |
|---|---|---|---|
| ΔV | 20mV | N | 256 |
| ($I_{evl}$) | 60μA | M, K : | |
| $I_{pre}$ | 120μA | ADC | 8, 8 |
| Tevl | 12ns | Edge det. | 3, 1 |
| Tpre | 2ns | halfton | 4, 1 |
| $C_{BL}$ | 200fF | $C_{BL}$ | 200fF |
| $V_{read}$ | 100mV | α | 0.5 |

Table-III
Magnet-Parameters

| $Ku_2$ (biaxial anisotropy) | $2 \times 10^6$ erg/cm$^3$ | polarization constant | High: 0.9 Low: 0.1 |
|---|---|---|---|
| Magnet Size (nm$^3$) | neuron | 60x20x1 | Damping coefficient | 0.007 |
| | DWM | 350x80x10 | Channel material | Cu |
| $H_k$ (coercively) | 5KOe | Channel spin flip length | 1μm |
| Ms (saturation magnetization) | 500emu/cm$^3$ | resistivity | 7Ω-nm |

Table-IV
Comparison with CMOS designs for feature extraction

| | CMOS Tech (T) | Fps (frame rate) | N (# PE) | Power | FOM* | FOM(proposed)/ FOM (given) |
|---|---|---|---|---|---|---|
| [45] | 0.35μ | 2000 | 32x32 | 600μW | $3.4 \times 10^3$ | 253 |
| [4] | 0.6μ | 100k | 1x1 | 85μW (per PE) | $1.1 \times 10^3$ | 200 |
| [31] | 0.25μ | 4000 | 128x128 | 20mW | $3.2 \times 10^3$ | 470 |
| [46] | 0.35μ | 2000 | 160x120 | 25mW | $1.5 \times 10^3$ | 560 |
| [47] | 0.35μ | 100 | 1 | 0.06μw | $1.66 \times 10^3$ | 500 |

Table-V
Comparison of the proposed ADC with state of art CMOS design

| Ref 8 bit | CMOS tech. | Fs | Power (W) | Spintronic ADC (W) | FOM** ratio |
|---|---|---|---|---|---|
| [35] | 0.18μ | 370KHz | 32 μ | 0.04μ | 133 |
| [36] | 0.18μ | 500kh | 7.75μ | 0.06μ | 32 |
| [37] | 0.25μ | 100KHz | 31μ | 0.012μ | 40 |
| [38] | 90nm | 10M | 70μ | 1μ | 70 |
| [39] | 90nm | 20Mhz | 290μ | 4μ | 72 |

*FOM = ($S^2$) x(#PE x Fps )/Power    **FOM = ($S^2$) /Power    S : technology scaling ratio

In this work we have assumed two supply sources $Vdd$ and $Vdd$- $\Delta V$. It can be assumed that charge supplied by the higher supply, is restored in the second source, and, can be utilized by other circuit components in a large-scale, heterogenous architecture. Effect of supply noise needs a more thorough analysis. Supply routing techniques, that can exploit the differential supply scheme employed in this work to mitigate the effects of supply noise, need to be explored.

Though, high precision computation on analog images may seem challenging with the technology limits associated with supply noise, the proposed scheme can be highly suitable for several low-level and middle-level analog image processing applications, for which, the conventional mixed signal designs consume large amount of power.

## VII. CONCLUSION

In this work we explored the application of the proposed spintronic neuron, in on-sensor image processing applications. It was shown that a spin-CMOS hybrid PE can handle analog processing functionality in an highly energy-efficient manner. The theoritical analysis presented, showed that, substituting some of the conventional analog processing units in an image acquision and processing hardware, by the spintronic neuron, can achieve ultra low power computation. This can facilitate the design of very high integration density hardware for sensory signal acquisition and processing.

## ACKNOWLEDGEMENT

This research was funded in part by Nano Research Initiative and by the INDEX center.